\documentclass[oneside]{article}
\usepackage[T1]{fontenc}
\usepackage{authblk}
\usepackage{float}
\usepackage{amsmath}
\usepackage{graphicx}
\usepackage{xfrac}
\usepackage[english]{babel}
\usepackage{lmodern}
\usepackage[T1]{fontenc}
\usepackage[latin9]{inputenc}

\usepackage{csquotes}

\usepackage{mathtools}
\usepackage{graphicx}
\usepackage{esint}
\usepackage[unicode=true,
 bookmarks=false,
breaklinks=false,pdfborder={0 0 1},backref=false,colorlinks=false]
 {hyperref}
\usepackage[a4paper, total={210mm,297mm},margin=2.0cm]{geometry}

\title{A multi-component discrete Boltzmann model for nonequilibrium reactive flows}

\author[1]{Chuandong Lin \thanks{Electronic address: \texttt{chuandonglin@163.com}; Corresponding author}}
\author[1,2]{Kai Hong Luo \thanks{Electronic address: \texttt{K.Luo@ucl.ac.uk}; Corresponding author}}
\author[1]{Linlin Fei}
\author[3]{Sauro Succi}

\affil[1]{Center for Combustion Energy, Key Laboratory for Thermal Science and Power Engineering of Ministry of Education, Department of Thermal Engineering, Tsinghua University, Beijing 100084, China}
\affil[2]{Department of Mechanical Engineering, University College London, Torrington Place, London WC1E 7JE, UK}
\affil[3]{Istituto Applicazioni Calcolo, CNR, Via dei Taurini 19, 00185 Rome, Italy}

\date{\today}

\begin{document}

\maketitle

\begin{abstract}
We propose a multi-component discrete Boltzmann model (DBM) for premixed, nonpremixed, or partially premixed nonequilibrium reactive flows. This model is suitable for both subsonic and supersonic flows with or without chemical reaction and/or external force. A two-dimensional sixteen-velocity model is constructed for the DBM. In the hydrodynamic limit, the DBM recovers the modified Navier-Stokes equations for reacting species in a force field. Compared to standard lattice Boltzmann models, the DBM presents not only more accurate hydrodynamic quantities, but also detailed nonequilibrium effects that are essential yet long-neglected by traditional fluid dynamics. Apart from nonequilibrium terms (viscous stress and heat flux) in conventional models, specific hydrodynamic and thermodynamic nonequilibrium quantities (high order kinetic moments and their departure from equilibrium) are dynamically obtained from the DBM in a straightforward way. Due to its generality, the developed methodology is applicable to a wide range of phenomena across many energy technologies, emissions reduction, environmental protection, mining accident prevention, chemical and process industry. 
\end{abstract}

\section*{Introduction}

Reactive flows are ubiquitous in nature and paramount to the sustainable development of society and ecological environment all over the world. For example, chemical energy released from fossil fuel in combustion comprises over $80\%$ the world's energy utilization \cite{Chu2012}. As the main human's threats, atmospheric pollution, climate change and global warming are directly relevant to harmful emissions from reactive flows, which involve a broad range of physicochemical phenomena, interacting over various spatial and temporal scales \cite{Law2006}. Besides, understanding of reactive flows is helpful to prevent fires in buildings, gas explosion in mines, burst in chemical factories, etc. Due to their significant importance to human society, reactive flows have attracted considerable attention in experimental, theoretical, and numerical fields. Actually, it is a challenging issue for traditional macroscopic or microscopic models to efficiently and accurately describe combustion phenomena where the span of spatial-temporal scales is relatively large and nonequilibrium phenomena play essential roles \cite{Liu2017,Ju-Review2014,Nagnibeda2009}. In fact, the nonequilibrium effects always change physical quantities (such as density, velocity, temperature, etc) in the evolution of fluid systems away from equilibrium, especially in transient and/or extreme conditions. A promising way to address this issue is to employ a mesoscopic kinetic model, lattice Boltzmann model (LBM), based on suitably simplified versions of the Boltzmann equation \cite{Succi-Book,Sofonea2001PA,Montessori2016,Qin2005,Zhang2008,Zhang2011,He1997,Benzi2009JCP,LiQing2016}. 

Recently, LBM has emerged as a versatile tool to simulate various complex systems, including reactive flows \cite{Ponce1993,Zanette1994,Qian1995,Weimar1996,Tian2016,Succi1997,Filippova2000CPC,Yu2002,Yamamoto2005,Lee2006,Chiavazzo2011,ChenSheng2012,succi2001applying,Yeomans2006,Ashna2017,Falcucci2016MN,Falcucci2017CES}. Previous LBMs were successfully employed as solvers of macroscopic governing equations, such as hydro-chemical equations for incompressible systems with low Mach number \cite{Ponce1993,Zanette1994,Qian1995,Weimar1996,Tian2016,Succi1997,Filippova2000CPC,Yu2002,Yamamoto2005,Lee2006,Chiavazzo2011,ChenSheng2012,succi2001applying,Yeomans2006,Ashna2017,Falcucci2016MN,Falcucci2017CES}. Physical quantities (such as pressure, velocity, temperature) can be described separately by several distribution functions in traditional LBMs, which are different from the Boltzmann equation where a single distribution function contains all information. For traditional LBMs, only a few low order kinetic moments of discrete equilibrium function are used, and the high order moments are not correctly reproduced \cite{Kang2014}. This limitation results in the failure of recovering the complete Navier-Stokes (NS) equations and providing more information on nonequilibrium behaviours. To overcome those problems, one promising method is to resort to a variant of traditional LBM, discrete Boltzmann model (DBM), where a required number of high order moments are satisfied \cite{XuGan2015SM,XuLai2016,XuYan2013,Lin2014CTP,Lin2015PRE,Lin2016CNF}. Different from traditional LBMs, DBM contains both equilibrium and nonequilibrium physical quantities that stem from the same discrete distribution function \cite{XuGan2015SM,XuLai2016,XuYan2013,Lin2014CTP,Lin2015PRE,Lin2016CNF}. 

Over the past years, the versatile DBM has been effectively applied to thermal phase separation, fluid instabilities, reactive flows, etc. \cite{XuGan2015SM,XuLai2016,XuYan2013,Lin2014CTP,Lin2015PRE,Lin2016CNF} The DBM for reactive flows was firstly presented by Yan et al. in 2013 \cite{XuYan2013}. Then, Lin et al. extended the DBM to reactive flows in a polar coordinate \cite{Lin2014CTP}. In 2015, Xu et al. proposed a multiple-relaxation-time DBM for reactive flows where the specific heat ratio and Prandtl number are adjustable \cite{Lin2015PRE}. The next year, a DBM is formulated for reactive flows where chemical reactant and product are described by two coupled distribution functions \cite{Lin2016CNF}. However, previous DBMs are suitable for premixed reactive flows, but not for nonpremixed or partially premixed reactive flows \cite{XuYan2013,Lin2014CTP,Lin2015PRE,Lin2016CNF}. For the sake of simulating both subsonic and supersonic nonequilibrium reactive flows with premixed, nonpremixed, or partially premixed reactants, we propose a multi-component DBM in this work. The DBM presents two ways to access the thermodynamic nonequilibrium behaviours. One is to measure the viscous stress and heat flux that are described by traditional NS models; The other is to calculate the kinetic moments of the difference between equilibrium and nonequilibrium discrete distribution functions, which is beyond conventional hydrodynamic models. Such capability is the main object of the present work.

\section*{Discrete Boltzmann model}

Without loss of generality, we consider the oxidation of propane in air using the one-step overall reaction, 
\begin{equation}
	{{\text{C}}_{3}}{{\text{H}}_{8}}+5{{\text{O}}_{2}}\to 3\text{C}{{\text{O}}_{2}}+4{{\text{H}}_{2}}\text{O}
	\label{ReactionFunction}
	\tt{,}
\end{equation}
where ${{\text{C}}_{3}}{{\text{H}}_{8}}$, ${{\text{O}}_{2}}$, ${\text{C}{{\text{O}}_{2}}}$, and ${{\text{H}}_{2}}\text{O}$ denote propane, oxygen, carbon dioxide, and water, respectively. The stoichiometric coefficients for them are $[a^{{{\text{C}}_{3}}{{\text{H}}_{8}}}, a^{{{\text{O}}_{2}}}, a^{{\text{C}{{\text{O}}_{2}}}}, a^{{{\text{H}}_{2}}\text{O}}]=[-1, -5, 3, 4]$. Nitrogen is assumed to be inert. The overall reaction rate reads
\begin{equation}
	\omega_{\text{ov}}={{k}_{\text{ov}}}{{n}^{{{\text{C}}_{3}}{{\text{H}}_{8}}}}{{n}^{{{\text{O}}_{2}}}}\exp \left( -{{E}_{a}}/RT \right)
	\tt{,}
\end{equation}
with $k_{ov}$ the reaction coefficient, $n^{\sigma}$ molar concentration, $E_a$ effective activation energy, $R$ universal gas constant, $T$ temperature. The mass change rate of species ${\sigma}$ is ${{\omega }^{\sigma }}={{a}^{\sigma }}\cdot {{\text{M}}^{\sigma }}\cdot {{\omega }_{\text{ov}}}$. In addition to the one-step reaction, detailed or reduced multi-step chemical kinetics, can also be employed. 

\begin{figure}[tbp]
	\begin{center}
		\includegraphics[width=0.35\textwidth]{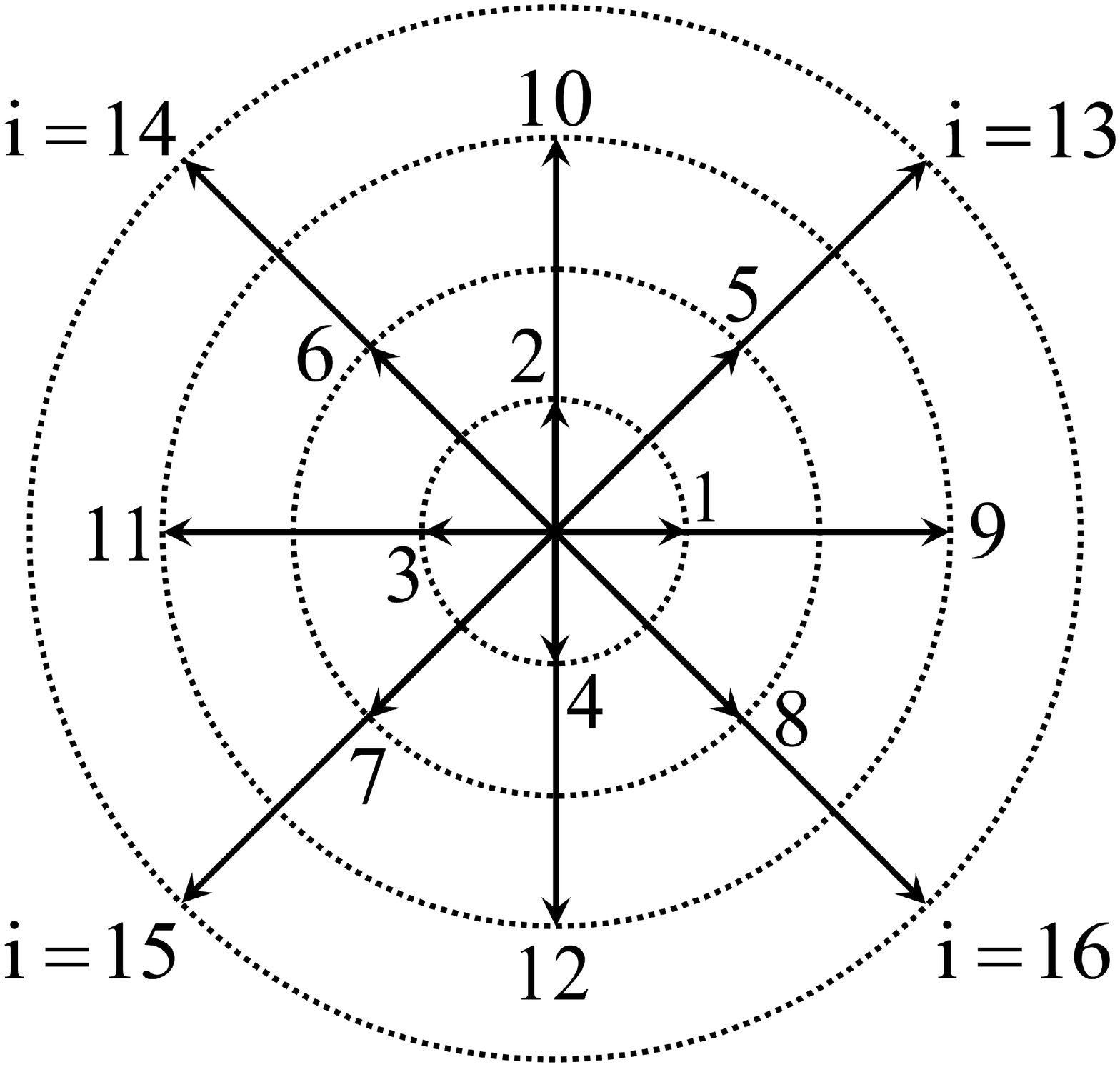}
	\end{center}
	\caption{Sketch of the two-dimensional sixteen-velocity model.}
	\label{Fig01}
\end{figure}
Let us introduce the discrete Boltzmann equation,
\begin{equation}
\frac{\partial f_{i}^{\sigma }}{\partial t}+\mathbf{v}_{i}^{\sigma }\cdot \nabla f_{i}^{\sigma }=\Omega _{i}^{\sigma }+R_{i}^{\sigma }+G_{i}^{\sigma }
\label{DiscreteBoltzmannEquation}
\tt{,}
\end{equation}
with
\begin{equation}
\Omega _{i}^{\sigma }=-\frac{1}{{{\tau }^{\sigma }}}\left [ f_{i}^{\sigma }-f_{i}^{\sigma eq}\left( {{n}^{\sigma }}, \mathbf{u}, T \right) \right]
\label{CollisionTerm}
\tt{,}
\end{equation}
\begin{equation}
R_{i}^{\sigma }=\frac{1}{{{\tau }^{\sigma }}}\left[ f_{i}^{\sigma eq}\left( {{n}^{\sigma *}}, \mathbf{u}, {{T}^{*}} \right)-f_{i}^{\sigma eq}\left( {{n}^{\sigma }}, \mathbf{u}, T \right) \right]
\label{ReactionTerm}
\tt{,}
\end{equation}
\begin{equation}
G_{i}^{\sigma }=\frac{1}{{{\tau }^{\sigma }}}\left[ f_{i}^{\sigma eq}\left( {{n}^{\sigma }}, {\mathbf{u}}^{\dagger \sigma }, {T}^{\dagger \sigma } \right)-f_{i}^{\sigma eq}\left( {{n}^{\sigma }}, {{\mathbf{u}}^{\sigma }}, {T}^{\sigma } \right) \right]
\label{ForceTerm}
\tt{.}
\end{equation}
Here $f_{i}^{\sigma}$ ($f_{i}^{\sigma eq}$) indicates the discrete (equilibrium) distribution function, $\mathbf{v}_i^{\sigma}$ the discrete velocity, $t$ (${\tau }^{\sigma}$) the (relaxation) time. $\Omega _{i}^{\sigma }$, $R_{i}^{\sigma }$, and $G_{i}^{\sigma}$ are the collision, reaction and force terms accounting for the molecular collision, chemical reaction and external force, respectively. 
The collision term in Eq. (\ref{CollisionTerm}) obeys the conversion of mass, momentum, and energy, from which the relations between the physical quantities (${n}^{\sigma }$, ${n}$, ${\mathbf{u}}^{\sigma }$, ${\mathbf{u}}$, ${T}^{\sigma }$, ${T}$) and the distribution function $f_{i}^{\sigma }$ are obtained \cite{Lin2016CNF}. The symbols with (without) superscript $\sigma$ denote the physical quantities of the species (mixture).
In Eq. (\ref{ReactionTerm}), ${n}^{\sigma *}$ and  ${T}^{*}$ (${n}^{\sigma}$ and  $T$) denote the molar concentration and temperature after (before) chemical reaction within time step $\tau^{\sigma}$. 
Similarly, in Eq. (\ref{ForceTerm}), the hydrodynamic velocity changes from ${\mathbf{u}}^{\sigma }$ to ${\mathbf{u}}^{\dagger \sigma }$ within time $\tau^{\sigma}$  due to external force, meanwhile the temperature changes from ${T}^{\sigma }$ to ${T}^{\dagger \sigma }$. The discrete equilibrium function $f_{i}^{\sigma eq}$ is linked with the formula,
\begin{equation}
{{f}^{\sigma eq}}={{n}^{\sigma }}{{\left( \frac{{{m}^{\sigma }}}{2\pi  T} \right)}^{D/2}}{{\left( \frac{{{m}^{\sigma }}}{2\pi {{I}^{\sigma }} T} \right)}^{1/2}}\exp \left[ -\frac{{{m}^{\sigma }}{{\left( \mathbf{v}-\mathbf{u} \right)}^{2}}}{2 T}-\frac{{{m}^{\sigma }}{{\eta }^{2}}}{2{{I}^{\sigma }} T} \right]
\label{FormulaFeq}
\tt{,}
\end{equation}
in the way that a required number of kinetic moments calculated by the integral of $f^{\sigma eq}$ are equivalent to those by the summation of $f^{\sigma eq}_i$. In Eq. (\ref{FormulaFeq}), $m^{\sigma}$ stands for molar mass, $D=2$ space dimension, $I^{\sigma}$ extra degrees of freedom corresponding to molecular rotation or vibration. There are $7$ moments satisfied by $f_{i}^{\sigma eq}=f_{i}^{\sigma eq}\left( {{n}^{\sigma }}, \mathbf{u}, T \right)$ in this work. Specifically, 
\begin{equation}
	{{n}^{\sigma }}=\sum\nolimits_{i}{f_{i}^{\sigma eq}}
	\label{relation1}
	\tt{,} 
\end{equation}
\begin{equation}
	{{n}^{\sigma }}\mathbf{u}=\sum\nolimits_{i}{f_{i}^{\sigma eq}\mathbf{v}_{i}^{\sigma }}
	\tt{,} 
\end{equation}
\begin{equation}
	{{n}^{\sigma }}\left[ \left( D+{{I}^{\sigma }} \right)\frac{T}{{{m}^{\sigma }}}+\mathbf{u}\cdot \mathbf{u} \right]=\sum\nolimits_{i}{f_{i}^{\sigma eq}\left( \mathbf{v}_{i}^{\sigma }\cdot \mathbf{v}_{i}^{\sigma }+\eta _{i}^{\sigma 2} \right)}
	\label{relation3}
	\tt{,} 
\end{equation}
\begin{equation}
	{{n}^{\sigma }}\left( {{\delta }_{\alpha \beta }}\frac{T}{{{m}^{\sigma }}}{{\mathbf{e}}_{\alpha }}{{\mathbf{e}}_{\beta }}+\mathbf{uu} \right)=\sum\nolimits_{i}{f_{i}^{\sigma eq}\mathbf{v}_{i}^{\sigma }\mathbf{v}_{i}^{\sigma }}
	\label{relation4}
	\tt{,} 
\end{equation}
\begin{equation}
	{{n}^{\sigma }}\mathbf{u}\left[ \left( D+{{I}^{\sigma }}+2 \right)\frac{T}{{{m}^{\sigma }}}+\mathbf{u}\cdot \mathbf{u} \right]=\sum\nolimits_{i}{f_{i}^{\sigma eq}\left( \mathbf{v}_{i}^{\sigma }\cdot \mathbf{v}_{i}^{\sigma }+\eta _{i}^{\sigma 2} \right)\mathbf{v}_{i}^{\sigma }}
	\tt{,} 
\end{equation}
\begin{equation}
	{{n}^{\sigma }}\left( {{\mathbf{u}}_{\alpha }}{{\mathbf{e}}_{\beta }}{{\mathbf{e}}_{\chi }}{{\delta }_{\beta \chi }}+{{\mathbf{e}}_{\alpha }}{{\mathbf{u}}_{\beta }}{{\mathbf{e}}_{\chi }}{{\delta }_{\alpha \chi }}+{{\mathbf{e}}_{\alpha }}{{\mathbf{e}}_{\beta }}{{\mathbf{u}}_{\chi }}{{\delta }_{\alpha \beta }} \right)\frac{{{T}^{\sigma }}}{{{m}^{\sigma }}}+{{n}^{\sigma }}\mathbf{uuu}=\sum\nolimits_{i}{f_{i}^{\sigma eq}\mathbf{v}_{i}^{\sigma }\mathbf{v}_{i}^{\sigma }\mathbf{v}_{i}^{\sigma }}
	\tt{,} 
\end{equation}
\begin{align}
	& {{n}^{\sigma }}{{\delta }_{\alpha \beta }}{{\mathbf{e}}_{\alpha }}{{\mathbf{e}}_{\beta }}\left[ \left( D+{{I}^{\sigma }}+2 \right)\frac{T}{{{m}^{\sigma }}}+\mathbf{u}\cdot \mathbf{u} \right]\frac{T}{{{m}^{\sigma }}}+{{n}^{\sigma }}\mathbf{uu}\left[ \left( D+{{I}^{\sigma }}+4 \right)\frac{T}{{{m}^{\sigma }}}+\mathbf{u}\cdot \mathbf{u} \right] \nonumber \\ 
	& =\sum\nolimits_{i}{f_{i}^{\sigma eq}\left( \mathbf{v}_{i}^{\sigma }\cdot \mathbf{v}_{i}^{\sigma }+\eta _{i}^{\sigma 2} \right)\mathbf{v}_{i}^{\sigma }\mathbf{v}_{i}^{\sigma }} 
	\label{relation7}
	\tt{,} 
\end{align}
which can be expressed in an uniform form ${\mathbf{\hat{f}}}^{eq}=\mathbf{M} {\mathbf{f}}^{eq}$, where 
\begin{equation}
	{\mathbf{f}}^{eq}={{\left( f_{1}^{\sigma eq}, f_{2}^{\sigma eq}, \cdots, f_{N_i}^{\sigma eq} \right)}^{\text{T}}}
	\nonumber
	\tt{,}
\end{equation}
\begin{equation}
	{\mathbf{\hat{f}}}^{eq}={{\left( \hat{f}_{1}^{\sigma eq}, \hat{f}_{2}^{\sigma eq}, \cdots, \hat{f}_{N_i}^{\sigma eq} \right)}^{\text{T}}}
	\nonumber
	\tt{,}
\end{equation}
\begin{equation}
	\mathbf{M}=\left( \begin{matrix}
		M_{11}^{\sigma } & M_{12}^{\sigma } & \cdots  & M_{1{{N}_{i}}}^{\sigma }  \\
		M_{21}^{\sigma } & M_{22}^{\sigma } & \cdots  & M_{2{{N}_{i}}}^{\sigma }  \\
		\vdots  & \vdots  & \ddots  & \vdots   \\
		M_{{{N}_{i}}1}^{\sigma } & M_{{{N}_{i}}2}^{\sigma } & \cdots  & M_{{{N}_{i}}{{N}_{i}}}^{\sigma }  \\
	\end{matrix} \right)
	\nonumber
	\tt{,}
\end{equation}
with $N_i=16$. The elements of ${\mathbf{\hat{f}}}^{eq}$ are
$\hat{f}_{1}^{\sigma eq}={{n}^{\sigma }}$, $\hat{f}_{2}^{\sigma eq}={{n}^{\sigma }}{{u}_{x}}$, $\hat{f}_{3}^{\sigma eq}={{n}^{\sigma }}{{u}_{y}}$, $\hat{f}_{4}^{\sigma eq}={{n}^{\sigma }}[ ( D+{{I}^{\sigma }} ){T/{{m}^{\sigma }}}+{{u}^{2}} ]$,
$\hat{f}_{5}^{\sigma eq}={{n}^{\sigma }}( {T/{{m}^{\sigma }}}+u_{x}^{2} )$, $\hat{f}_{6}^{\sigma eq}={{n}^{\sigma }}{{u}_{x}}{{u}_{y}}$, $\hat{f}_{7}^{\sigma eq}={{n}^{\sigma }}( {T/{{m}^{\sigma }}}+u_{y}^{2} )$, $\hat{f}_{8}^{\sigma eq}={{n}^{\sigma }}{{u}_{x}}[ ( D+{{I}^{\sigma }}+2 ){T/{{m}^{\sigma }}}+{{u}^{2}} ]$, $\hat{f}_{9}^{\sigma eq}={{n}^{\sigma }}{{u}_{y}}[ ( D+{{I}^{\sigma }}+2 ){T/{{m}^{\sigma }}}+{{u}^{2}} ]$, $\hat{f}_{10}^{\sigma eq}=3{{n}^{\sigma }}{{u}_{x}}{T/{{m}^{\sigma }}}+{{n}^{\sigma }}u_{x}^{3}$, $\hat{f}_{11}^{\sigma eq}={{n}^{\sigma }}{{u}_{y}}{T/{{m}^{\sigma }}}+{{n}^{\sigma }}u_{x}^{2}{{u}_{y}}$, $\hat{f}_{12}^{\sigma eq}={{n}^{\sigma }}{{u}_{x}}{T/{{m}^{\sigma }}}+{{n}^{\sigma }}{{u}_{x}}u_{y}^{2}$, $\hat{f}_{13}^{\sigma eq}=3{{n}^{\sigma }}{{u}_{y}}{T/{{m}^{\sigma }}}+{{n}^{\sigma }}u_{y}^{3}$, $\hat{f}_{14}^{\sigma eq}={{n}^{\sigma }}[ ( D+{{I}^{\sigma }}+2 ){T/{{m}^{\sigma }}}+{{u}^{2}} ]{T/{{m}^{\sigma }}}+{{n}^{\sigma }}u_{x}^{2}[ ( D+{{I}^{\sigma }}+4 ){T/{{m}^{\sigma }}}+{{u}^{2}} ]$,
$\hat{f}_{15}^{\sigma eq}={{n}^{\sigma }}{{u}_{x}}{{u}_{y}}[ ( D+{{I}^{\sigma }}+4 ){T/{{m}^{\sigma }}}+{{u}^{2}} ]$, $\hat{f}_{16}^{\sigma eq}={{n}^{\sigma }}[ ( D+{{I}^{\sigma }}+2 ){T/{{m}^{\sigma }}}+{{u}^{2}} ]{T/{{m}^{\sigma }}}+{{n}^{\sigma }}u_{y}^{2}[ ( D+{{I}^{\sigma }}+4 ){T/{{m}^{\sigma }}}+{{u}^{2}} ]$,
and those of $\mathbf{M}$ are
$M_{1i}^{\sigma }=1$, $M_{2i}^{\sigma }=v_{ix}^{\sigma }$, $M_{3i}^{\sigma }=v_{iy}^{\sigma }$, $M_{4i}^{\sigma }=v_{i}^{\sigma 2}+\eta _{i}^{\sigma 2}$, ${{M}_{5i}}=v_{ix}^{\sigma 2}$,
${{M}_{6i}}=v_{ix}^{\sigma }v_{iy}^{\sigma }$, ${{M}_{7i}}=v_{iy}^{\sigma 2}$, ${{M}_{8i}}=( v_{i}^{\sigma 2}+\eta _{i}^{\sigma 2} )v_{ix}^{\sigma }$, ${{M}_{9i}}=( v_{i}^{\sigma 2}+\eta _{i}^{\sigma 2} )v_{ix}^{\sigma }$,
${{M}_{10i}}=v_{ix}^{\sigma 3}$, ${{M}_{11i}}=v_{ix}^{\sigma 2}v_{iy}^{\sigma }$, ${{M}_{12i}}=v_{ix}^{\sigma }v_{iy}^{\sigma 2}$, ${{M}_{13i}}=v_{iy}^{\sigma 3}$,
${{M}_{14i}}=( v_{i}^{\sigma 2}+\eta _{i}^{\sigma 2} )v_{ix}^{\sigma 2}$, ${{M}_{15i}}=( v_{i}^{\sigma 2}+\eta _{i}^{\sigma 2} )v_{ix}^{\sigma }v_{iy}^{\sigma }$, ${{M}_{16i}}=( v_{i}^{\sigma 2}+\eta _{i}^{\sigma 2} )v_{iy}^{\sigma 2}$. 

The discrete velocities $\mathbf{v}_{i}^{\sigma }$ and $\eta _{i}^{\sigma }$ are (see Fig. \ref{Fig01}),
\begin{equation}
\left[ \mathbf{v}_{i}^{\sigma }, \eta _{i}^{\sigma } \right]=\left\{ \begin{array}{*{35}{l}}
\left[ \text{cyc}: v_{a}^{\sigma }\left( \pm 1,0 \right),{{\eta }_{a}^{\sigma }} \right] & 1\le i\le 4 \tt{,}  \\
\left[ \text{cyc}: v_{b}^{\sigma }\left( \pm 1,\pm 1 \right),{{\eta }_{b}^{\sigma }} \right] & 5\le i\le 8 \tt{,}  \\
\left[ \text{cyc}: v_{c}^{\sigma }\left( \pm 1,0 \right),{{\eta }_{c}^{\sigma }} \right] & 9\le i\le 12 \tt{,}  \\
\left[ \text{cyc}: v_{d}^{\sigma }\left( \pm 1,\pm 1 \right),{{\eta }_{d}^{\sigma }} \right] & 13\le i\le 16 \tt{,}
\end{array} \right.
\end{equation}
where cyc indicates the cyclic permutation, $(v_{a}^{\sigma }$, $v_{b}^{\sigma }$, $v_{c}^{\sigma }$, $v_{d}^{\sigma }$, $\eta_{a}^{\sigma }$, $\eta_{b}^{\sigma }$, $\eta_{c}^{\sigma }$, $\eta_{d}^{\sigma })$ are adjustable parameters. And the discrete equilibrium distribution is expressed by
\begin{equation}
{{\mathbf{f}}^{eq}}={{\mathbf{M}}^{-1}} {{\mathbf{\hat{f}}}^{eq}}
\tt{.}
\end{equation}

One significant capability of the DBM is to investigate nonequilibrium manifestations by measuring the following physical variables, 
\begin{equation}
\Delta_{2}^{\sigma }=\sum\nolimits_{i}{{{m}^{\sigma }}\left( f_{i}^{\sigma }-f_{i}^{\sigma eq} \right)\mathbf{v}_{i}^{\sigma}\mathbf{v}_{i}^{\sigma}}
\label{Delta2} \tt{,}
\end{equation}
\begin{equation}
\Delta_{3,1}^{\sigma }=\sum\nolimits_{i}{{{m}^{\sigma }}\left( f_{i}^{\sigma }-f_{i}^{\sigma eq} \right)\left( \mathbf{v}_{i}^{\sigma } \cdot \mathbf{v}_{i}^{\sigma }+\eta _{i}^{\sigma 2} \right)\mathbf{v}_{i}^{\sigma }}
\label{Delta31} \tt{,}
\end{equation}
\begin{equation}
\Delta_{3}^{\sigma }=\sum\nolimits_{i}{{{m}^{\sigma }}\left( f_{i}^{\sigma }-f_{i}^{\sigma eq} \right)\mathbf{v}_{i}^{\sigma}\mathbf{v}_{i}^{\sigma}\mathbf{v}_{i}^{\sigma}}
\label{Delta3} \tt{,}
\end{equation}
\begin{equation}
\Delta_{4,2}^{\sigma }=\sum\nolimits_{i}{{{m}^{\sigma }}\left( f_{i}^{\sigma }-f_{i}^{\sigma eq} \right)\left( \mathbf{v}_{i}^{\sigma } \cdot \mathbf{v}_{i}^{\sigma }+\eta _{i}^{\sigma 2} \right)\mathbf{v}_{i}^{\sigma}\mathbf{v}_{i}^{\sigma}}
\label{Delta42} \tt{.}
\end{equation}
Mathematically, $\Delta_{2}^{\sigma }=\Delta_{2 \alpha \beta}^{\sigma } \mathbf{e}_{\alpha} \mathbf{e}_{\beta}$ is a second order tensor with four components, among which only three ($\Delta_{2 xx}^{\sigma }$, $\Delta_{2 xy}^{\sigma }$, $\Delta_{2 yy}^{\sigma }$) are independent;  $\Delta_{3}^{\sigma }=\Delta_{3 \alpha \beta \gamma}^{\sigma } \mathbf{e}_{\alpha} \mathbf{e}_{\beta} \mathbf{e}_{\gamma}$ is a third order tensor with eight components where only four ($\Delta_{3 xxx}^{\sigma }$, $\Delta_{3 xxy}^{\sigma }$, $\Delta_{3 xyy}^{\sigma }$, $\Delta_{3 yyy}^{\sigma }$) are independent; $\Delta_{3,1}^{\sigma }=\Delta_{3,1 \alpha}^{\sigma } \mathbf{e}_{\alpha}$ is the first order tensor (i.e. vector) contracted from a third order tensor and have two independent components; Similarly for $\Delta_{4,2}^{\sigma }=\Delta_{4,2 \alpha \beta}^{\sigma } \mathbf{e}_{\alpha} \mathbf{e}_{\beta}$. 

It is easy to prove that, via the Chapman-Enskog multiscale analysis, the DBM is in line with the following modified NS equations,
\begin{equation}
\frac{\partial {{\rho }^{\sigma }}}{\partial t}+\frac{\partial }{\partial {{r}_{\alpha }}}\left( {{\rho }^{\sigma }}u_{\alpha }^{\sigma } \right)={{\rho }^{\sigma \prime }}
\label{NS1}
\tt{,}
\end{equation}
\begin{equation}
\frac{\partial }{\partial t}\left( {{\rho }^{\sigma }}u_{\alpha }^{\sigma } \right)+\frac{\partial }{\partial {{r}_{\beta }}}\left( {{\delta }_{\alpha \beta }}{{p}^{\sigma }}+{{\rho }^{\sigma }}u_{\alpha }^{\sigma }u_{\beta }^{\sigma }+\Delta_{2 \alpha \beta}^{\sigma } \right)={{\rho }^{\sigma }}{{a}_{\alpha }}-\frac{{{\rho }^{\sigma }}}{{{\tau }^{\sigma }}}\left( u_{\alpha }^{\sigma }-{{u}_{\alpha }} \right)+{{\rho }^{\sigma \prime }}{{u}_{\alpha }}
\label{NS2}
\tt{,}
\end{equation}
\begin{align}
& \frac{\partial }{\partial t}{{\rho }^{\sigma }}\left( {{e}^{\sigma }}+\frac{1}{2}{{u}^{\sigma 2}} \right)+\frac{\partial }{\partial {{r}_{\alpha }}}\left[ {{\rho }^{\sigma }}u_{\alpha }^{\sigma }\left( {{e}^{\sigma }}+\frac{1}{2}{{u}^{\sigma 2}} \right)+{{p}^{\sigma }}u_{\alpha }^{\sigma }+\Delta_{3,1 \alpha}^{\sigma } \right]  \nonumber \\
& ={{\rho }^{\sigma }}u_{\alpha }^{\sigma }{{a}_{\alpha }}+{{\rho }^{\sigma \prime }}\left( \frac{D+{{I}^{\sigma }}}{2}\frac{T}{{{m}^{\sigma }}}+\frac{1}{2}{{u}^{2}} \right)+\frac{D+{{I}^{\sigma }}}{2{{m}^{\sigma }}}{{\rho }^{\sigma }}{T}'-\frac{{{\rho }^{\sigma }}}{{{\tau }^{\sigma }}}\left( \frac{D+{{I}^{\sigma }}}{2}\frac{{{T}^{\sigma }}-T}{{{m}^{\sigma }}}+\frac{{{u}^{\sigma 2}}-{{u}^{2}}}{2} \right) 
\label{NS3}
\tt{,}
\end{align}
in the hydrodynamic limit, with
$
\Delta_{2 \alpha \beta}^{\sigma }=P_{\alpha \beta }^{\sigma }+U_{\alpha \beta }^{\sigma } 
$, 
$
\Delta_{3,1 \alpha}^{\sigma }=-{{\kappa }^{\sigma }}\frac{\partial }{\partial {{r}_{\alpha }}} ( \frac{D+{{I}^{\sigma }}}{2}\frac{{{T}^{\sigma }}}{{{m}^{\sigma }}} )+u_{\beta }^{\sigma }P_{\alpha \beta }^{\sigma }-X_{\alpha }^{\sigma }-Z_{\alpha }^{\sigma }
$, 
$
P_{\alpha \beta }^{\sigma }=-{{\mu }^{\sigma }} ( \frac{\partial u_{\alpha }^{\sigma }}{\partial {{r}_{\beta }}}+\frac{\partial u_{\beta }^{\sigma }}{\partial {{r}_{\alpha }}}-\frac{2{{\delta }_{\alpha \beta }}}{D+{{I}^{\sigma }}}\frac{\partial u_{\chi }^{\sigma }}{\partial {{r}_{\chi }}} )
$,
$
U_{\alpha \beta }^{\sigma }=-( {{\rho }^{\sigma }}+{{\rho }^{\sigma \prime }}{{\tau }^{\sigma }} )( {{\delta }_{\alpha \beta }}\frac{{{u}^{\sigma 2}}+{{u}^{2}}-2u_{\chi }^{\sigma }{{u}_{\chi }}}{D+{{I}^{\sigma }}}+{{u}_{\alpha }}u_{\beta }^{\sigma }+u_{\alpha }^{\sigma }{{u}_{\beta }}-u_{\alpha }^{\sigma }u_{\beta }^{\sigma }-{{u}_{\alpha }}{{u}_{\beta }} )
$, 
$
X_{\alpha }^{\sigma }={{\tau }^{\sigma }}( D+{{I}^{\sigma }}+2 )\frac{{{\rho }^{\sigma }}\left( u_{\alpha }^{\sigma }-{{u}_{\alpha }} \right)}{{{m}_{\sigma }}}{T}'
$, 
$
Z_{\alpha }^{\sigma }=( {{\rho }^{\sigma }}+{{\tau }^{\sigma }}{{\rho }^{\sigma \prime }} )\{ \frac{u_{\alpha }^{\sigma }}{D+{{I}^{\sigma }}}{{( u_{\beta }^{\sigma }-{{u}_{\beta }} )}^{2}}-\frac{u_{\alpha }^{\sigma }-{{u}_{\alpha }}}{2}[ \frac{D+{{I}^{\sigma }}+2}{{{m}^{\sigma }}}( {{T}^{\sigma }}-T )+{{u}^{\sigma 2}}-{{u}^{2}} ] \}
$, 
where $p^{\sigma }=n^{\sigma}T^{\sigma }$ indicates pressure, $e^{\sigma }=(D+I^{\sigma})T^{\sigma }/(2m^{\sigma })$ internal energy per unit mass, $\mu ^{\sigma }=p^{\sigma }\tau ^{\sigma }$ dynamic viscosity coefficient, $\kappa^{\sigma}=\gamma^{\sigma} \mu ^{\sigma }$ heat conductivity, and $\gamma^{\sigma} =(D+I^{\sigma}+2)/(D+I^{\sigma})$ specific heat ratio. 
The superscript  ``$ ^{\prime}$" represents the change rate of physical quantities due to the chemical reaction. In fact, applying the operator $\sum_{\sigma }$ to both sides of Eqs. (\ref{NS1}) $-$ (\ref{NS3}) gives NS equations for the whole system, which reduces to conventional NS equations when $\textbf{u}^{\sigma}=\textbf{u}$ and $T^{\sigma}=T$. Obviously, Eqs. (\ref{NS1}) $-$ (\ref{NS3}) gives a more detailed description than the conventional NS equations. The latter is just a special case of the former. 

Furthermore, dynamic viscosity and heat conductivity in the NS equations are regarded as two important thermodynamic nonequilibrium manifestations or physical effects on fluid flows. In fact, a more detailed way to study the nonequilibrium behaviours is to investigate the departure of high order velocity moments from their local equilibrium counterparts, as shown in Eqs. (\ref{Delta2})-(\ref{Delta42}). Those kinetic moments of the difference between nonequilibrium and equilibrium distribution functions have significant physical meanings. In particular, $\Delta_{2}^{\sigma }$ is associated with viscous stress tensor and nonorganised momentum fluxes, $\Delta_{3,1}^{\sigma }$ and $\Delta_{3}^{\sigma }$ are related to nonorganised energy (heat) fluxes, $\Delta_{4,2}^{\sigma }$ corresponds to the flux of nonorganised energy (heat) flux \cite{Lin2014CTP,Lin2015PRE}. The terminology ``nonorganised" is relative to ``organised". The latter refers to the collective motion of a fluid flow, while the former corresponds to the molecular individualism on top of the collective motion \cite{XuLai2016}. Moreover, $\frac{1}{2} {\tt{M}}_{2 \alpha \alpha}(f^{\sigma })=\frac{1}{2} \sum\nolimits_{i}{{{m}^{\sigma }} f_{i}^{\sigma } {v}_{i \alpha}^{\sigma 2}}$ is defined as the translational energy of species $\sigma$ in $\alpha$ direction, $\frac{1}{2} {\tt{M}}_{2 \alpha \alpha}(f^{\sigma eq})$ is its equilibrium counterpart, and $\frac{1}{2} \Delta_{2 \alpha \alpha}^{\sigma }=\frac{1}{2} {\tt{M}}_{2 \alpha \alpha}(f^{\sigma })-\frac{1}{2} {\tt{M}}_{2 \alpha \alpha}(f^{\sigma eq})$ the nonorgnised energy of species $\sigma$ in $\alpha$ direction; $\frac{1}{2} {\tt{M}}_{3,1 \alpha}(f^{\sigma eq})=\frac{1}{2} \sum\nolimits_{i}{{{m}^{\sigma }} f_{i}^{\sigma eq} ({v}_{i}^{\sigma 2}+{\eta}_{i}^{\sigma 2}){v}_{i \alpha}^{\sigma}}$ refers to the organised flux of energy in $\alpha$ direction, $\frac{1}{2} {\Delta}_{3,1 \alpha}^{\sigma }$ the nonorganised flux of energy in $\alpha$ direction, and $\frac{1}{2} {\tt{M}}_{3,1 \alpha}(f^{\sigma })=\frac{1}{2} {\tt{M}}_{3,1 \alpha}(f^{\sigma eq})+\frac{1}{2} {\Delta}_{3,1 \alpha}^{\sigma }$ the total flux of energy in $\alpha$ direction. Obviously, DBM provides more nonequilibrium information on various species in fluid flows, which is an essential advantage over traditional models. 

\section*{Numerical simulation}\label{SecIII}

To validate this DBM, we conduct five simulation tests. Test one is the combustion of (premixed, nonpremixed, and partially premixed) propane-air filled in a free-falling box. The released heat in constant volume and the external force effects are demonstrated. The second test is a subsonic flame at constant pressure. In the third part, to show its suitability for high speed compressible systems, the DBM is used to simulate a shock wave. Its capability to investigate nonequilibrium effects is verified as well. A supersonic reacting wave is simulated in the fourth part. The first four tests are 1-dimensional (1-D) cases. The last is for a typical 2-D case, Kelvin-Helmholtz (KH) instability. 

Moreover, the first order Euler forward time discretization and the second order nonoscillatory and nonfree-parameters dissipative finite difference scheme \cite{Zhang1991NND} are adopted for the temporal and spatial derivatives in Eq. (\ref{DiscreteBoltzmannEquation}) in this section. Hence, the discrete velocities $\mathbf{v}_{i}$ are independent of the grid mesh $\Delta x$ and $\Delta y$. For the purpose of accuracy and robustness, it is preferable to set the values of discrete velocities ($v^{\sigma}_{a}$, $v^{\sigma}_{b}$, $v^{\sigma}_{c}$, $v^{\sigma}_{d}$) around the values of $|\mathbf{u}|$ and $\sqrt{D T/m^{\sigma}}$, and choose ($\eta^{\sigma}_{a}$, $\eta^{\sigma}_{b}$, $\eta^{\sigma}_{c}$, $\eta^{\sigma}_{d}$) around the value $\sqrt{I^{\sigma} T/m^{\sigma}}$, which is reasonable on account of Eqs. (\ref{relation1})-(\ref{relation3}). 

\subsection*{Combustion in constant volume}

\begin{figure}[tbp]
	\begin{center}
		\includegraphics[width=0.99\textwidth]{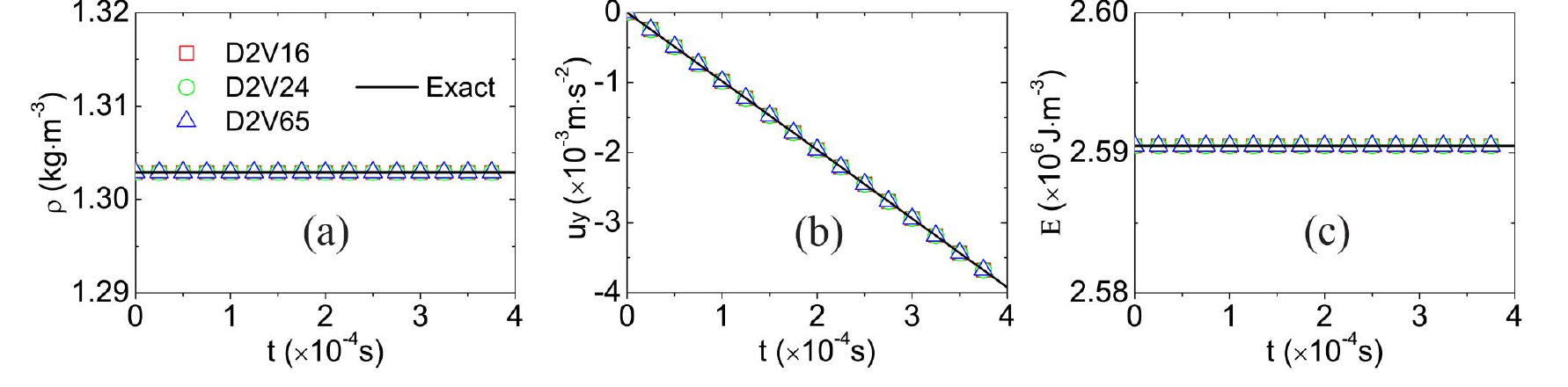}
	\end{center}
	\caption{The average physical quantities of the reactive mixture in the free-falling box versus time $t$: (a) mass density $\rho$, (b) vertical speed $u_y$, (c) the sum of internal energy and chemical heat $E$. The symbols (squares, circles, and triangles) denote simulation results (D2V16, D2V24, and D2V65), the solid lines exact solutions.}
	\label{Fig02}
\end{figure}

First of all, we simulate the combustion of propane-air filled in a free-falling box, which consists of three parts with volumes $V_1$, $V_2$ and $V_3$, respectively. The fixed volume of the box is $V_0=V_1+V_2+V_3$, and $V_1:V_2:V_3=3:119:78$. Initially, the left part is filled with propane, the middle part is full of air, and the right part is occupied by the propane-air mixture with equivalence ratio $0.6$. In each part, the particle number density is $40.6 \text{mol}\cdot {{\text{m}}^{-3}}$, temperature $300$ K, and pressure $1$ atm. Premixed, nonpremixed and partially premixed combustion phenomena take place simultaneously in this box after ignition. Specifically, the nonpremixed combustion takes place between the left and middle parts, the partially premixed combustion occurs between the middle and right parts with a changing equivalence ratio, and the premixed combustion is in the rightmost part with a constant equivalence ratio. Three discrete velocity models (D2V16, D2V24 \cite{Lin2015PRE}, and D2V65 \cite{Watari2007}) are employed for this simulation. The grid is $N_x \times N_y=200 \times 1$, spatial step $\Delta x= \Delta y =5 \times 10^{-7}$ m, temporal step $\Delta x= \Delta y =1.25 \times 10^{-10}$ s. 

Figure \ref{Fig02} illustrates the simulation results and exact solutions during the chemical reaction in the free-falling box. Theoretically, the density remains constant, $\rho=1.30290 \text{kg}\cdot {\text{m}}^{-3}$, the  velocity changes as ${u}_{y}=g t$, with $g=-9.8 \text{m} \cdot {\text{s}}^{-2}$, and the sum of internal energy and chemical heat remains constant, $E=2.59050\times 10^{6} \text{J} \cdot \text{m}^{-3}$. As for the simulation, each model (D2V16, D2V24 \cite{Lin2015PRE}, and D2V65 \cite{Watari2007}) gives the density $\rho=1.30290 \text{kg}\cdot {\text{m}}^{-3}$ and the energy $E=2.59050\times 10^{6} \text{J} \cdot \text{m}^{-3}$ in the whole process, which coincide with the exact solutions. There are tiny differences between the simulation results and exact solutions of the velocity. For example, at time $t=3 \times 10^{-4}$ s, the three models (D2V16, D2V24 \cite{Lin2015PRE}, and D2V65 \cite{Watari2007}) give simulation results $u_y=-2.9402 \times 10^{-3}$, $-2.9399 \times 10^{-3}$, and $-2.9401 \times 10^{-3}  \text{m} \cdot {\text{s}}^{-1}$, respectively. Compared to the exact value $u_y=-2.94 \times 10^{-3} \text{m} \cdot {\text{s}}^{-1}$, their relative errors are $0.0068\%$, $0.0034\%$, and $0.0034\%$, respectively. Obviously, all simulation results agree well with the exact solutions. 

Furthermore, after the chemical reaction is completed, the adiabatic constant volume temperature is $2078 \tt{K}$ calculated by the three DBMs, while it is $2614 \tt{K}$ obtained by the standard LBM \cite{He1997,Yamamoto2002}. The parameters for the LBM in this work are the same as those in Ref. \cite{Yamamoto2002}. Compared with the experimental datum $2080 \tt{K}$ \cite{Robert2016}, the relative differences are $0.1\%$ for the DBM and $25.7\%$ for the standard LBM, respectively. Physically, the DBM is suitable for compressible systems with adjustable ratio of specific heats, while the LBM in Refs. \cite{He1997,Yamamoto2002} can only be used for the case with constant pressure and fixed ratio of specific heats. 

\begin{center}
	\begin{table}[tbp]
		\caption{Computing times for simulation of combustion in constant volume with various models. }
		\begin{tabular}{c|c|c|c}
			\hline\hline
			~~Model~~ & ~~Number of discrete velocities~~ & ~~Number of moment relations~~ & ~~Computing time~~\\
			\hline
			D2V16 & $16$ & $16$ & $1560$ s \\
			\hline
			D2V24 & $24$ & $24$ & $3960$ s \\
			\hline
			D2V65 & $65$ & $16$ & $4980$ s \\
			\hline\hline
		\end{tabular}
		\label{TableI}
	\end{table}
\end{center}

To discuss computational costs of various discrete velocity models, we keep a record of computing times required by the aforementioned simulation in Table \ref{TableI}. The computational facility is a personal computer with Intel(R) Core(TM) i7-6700K CPU @ 4.00GHz and RAM 32.00 GB. There are $16$, $24$, and $65$ ($16$, $24$, and $16$) discrete velocities (moment relations) in D2V16, D2V24 \cite{Lin2015PRE}, and D2V65 \cite{Watari2007}, respectively. And the computing times are $1560$ s, $3960$ s, and $4980$ s for the three models, respectively. Obviously, D2V24 and D2V65 models need larger RAM and longer time than D2V16 model. 

\subsection*{Flame at constant pressure}\label{secIIIB}

\begin{figure}[tbp]
	\begin{center}
		\includegraphics[width=0.8\textwidth]{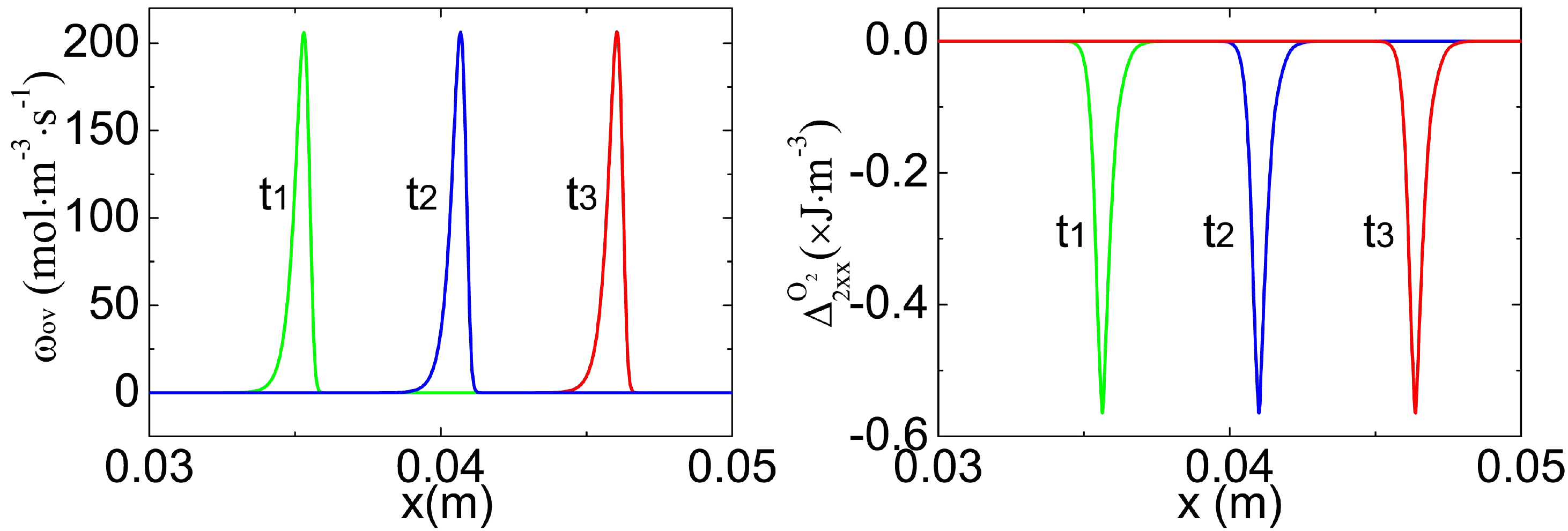}
	\end{center}
	\caption{The flame profiles at constant pressure at times $t_1=0.055 \tt{s}$, $t_2=0.0625 \tt{s}$, and $t_3=0.07 \tt{s}$. The left panel is for the overall reaction rate $\omega_{\tt{ov}}$, and the right for $\Delta_{2 xx}^{\tt{O}_2}$ which is two times the departure of translational energy of $\tt{O_2}$ in $x$ direction from its equilibrium counterpart.}
	\label{Fig03}
\end{figure}
\begin{figure}[tbp]
	\begin{center}
		\includegraphics[width=0.45\textwidth]{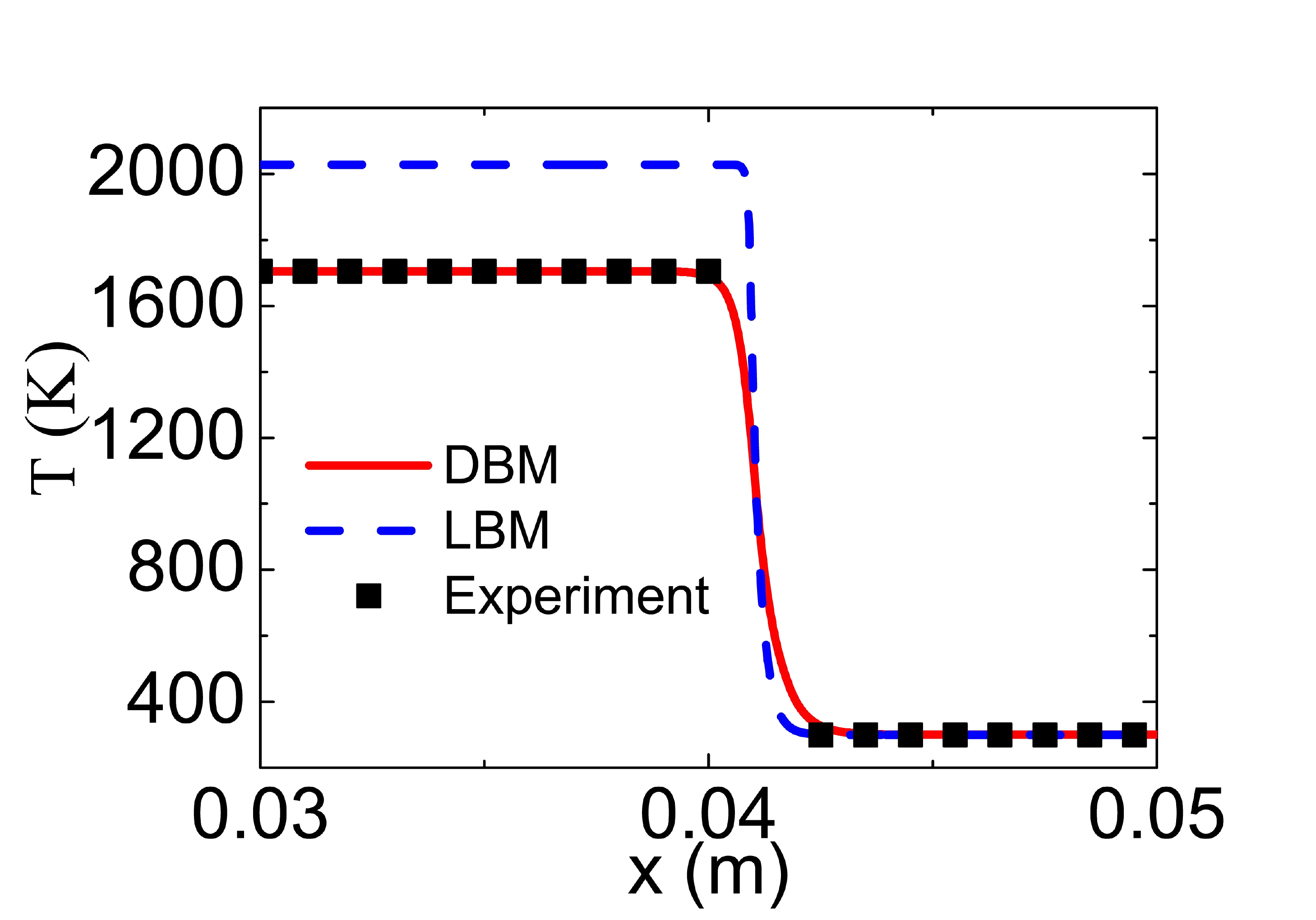}
	\end{center}
	\caption{The flame temperature at constant pressure. The solid (dashed) line denotes DBM (LBM) results, the squares experimental data.}
	\label{Fig04}
\end{figure}

Let us simulate a flame at constant pressure. It travels with subsonic speed in a channel from left to right. In front of the flame is the propane-air mixture with equivalence ratio $0.6$, particle number density $44.6 \text{mol}\cdot {{\text{m}}^{-3}}$, temperature $300$ K, and pressure $1$ atm. The grid is $N_x \times N_y=2500 \times 1$, spatial step $\Delta x= \Delta y =2 \times 10^{-5}$ m, temporal step $\Delta t =1.25 \times 10^{-10}$ s. 

Figure \ref{Fig03} shows the evolution of $\omega_{\tt{ov}}$ (left) and $\Delta_{2 xx}^{\tt{O}_2}$ (right) versus $x$. The peak of $\omega_{\tt{ov}}$ corresponding to the most active chemical reaction is ahead of the trough of $\Delta_{2 xx}^{\tt{O}_2}$ where nonequilibrium manifestations are intense and physical gradients are sharp. Note that the nonequilibrium manifestations can be employed to capture the flame or other interfaces \cite{XuLai2016}. The flame speed, $0.71 \tt{m/s}$, can be obtained from the profiles of either $\omega_{\tt{ov}}$ or $\Delta_{2 xx}^{\tt{O}_2}$. And the flow velocity is $0.60 \tt{m/s}$ in front of the flame. Hence, the burning velocity is $(0.71-0.60) \tt{m/s}=0.11 \tt{m/s}$, which equals the experimental result $0.11 \tt{m/s}$ \cite{Yamaoka1985}. While the standard LBM \cite{Yamamoto2002} gives a relative error, $9.1\%$, compared with the experimental result \cite{Yamaoka1985}. 

Moreover, in the DBM simulation, the pressure is close to $1\tt{atm}$ around the flame, and the temperature is $1705\tt{K}$ behind the flame, which is consistent with the experiment \cite{Law2006}, while the temperature is $2028\tt{K}$ in the traditional LBM \cite{He1997,Yamamoto2002} (see Fig. \ref{Fig04}). The latter's relative error is $18.9\%$ compared with the experimental result \cite{Law2006}. Physically, the ratio of specific heats in the DBM is tunable, while the one in the LBM in Refs. \cite{He1997,Yamamoto2002} is fixed at $2$. Besides, the chemical reaction does not affect the flow field in this LBM \cite{Yamamoto2002}, while the chemical reaction and fluid flow are naturally coupled in our DBM.

\subsection*{Shock wave}

\begin{figure}[tbp]
	\begin{center}
		\includegraphics[width=0.8\textwidth]{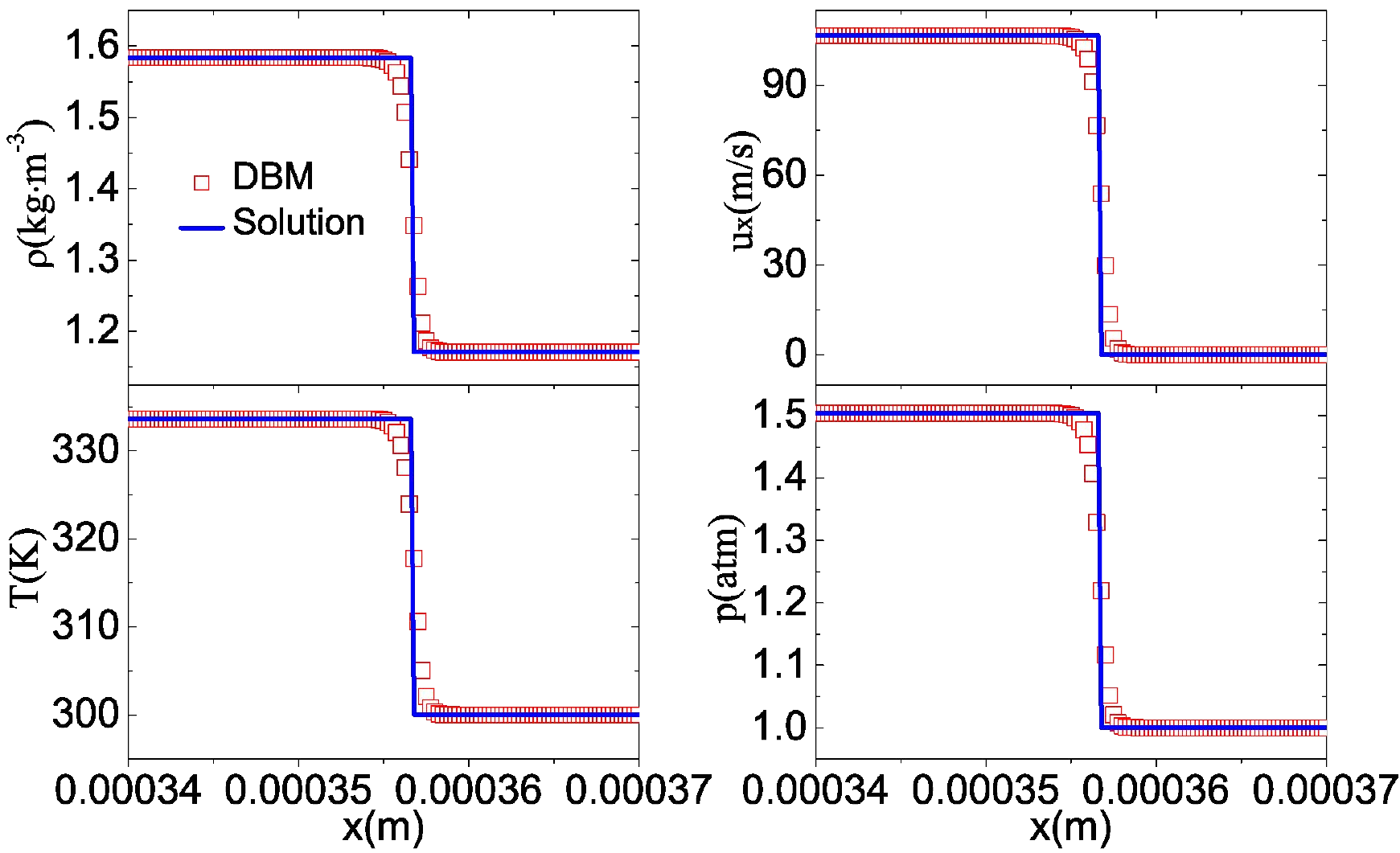}
	\end{center}
	\caption{The profiles of a shock wave: (a) mass density $\rho$, (b) velocity $u_x$, (c) temperature $T$, (d) pressure $p$. The squares represent DBM results, the lines exact solutions.}
	\label{Fig05}
\end{figure}
\begin{figure}[tbp]
	\begin{center}
		\includegraphics[width=0.8\textwidth]{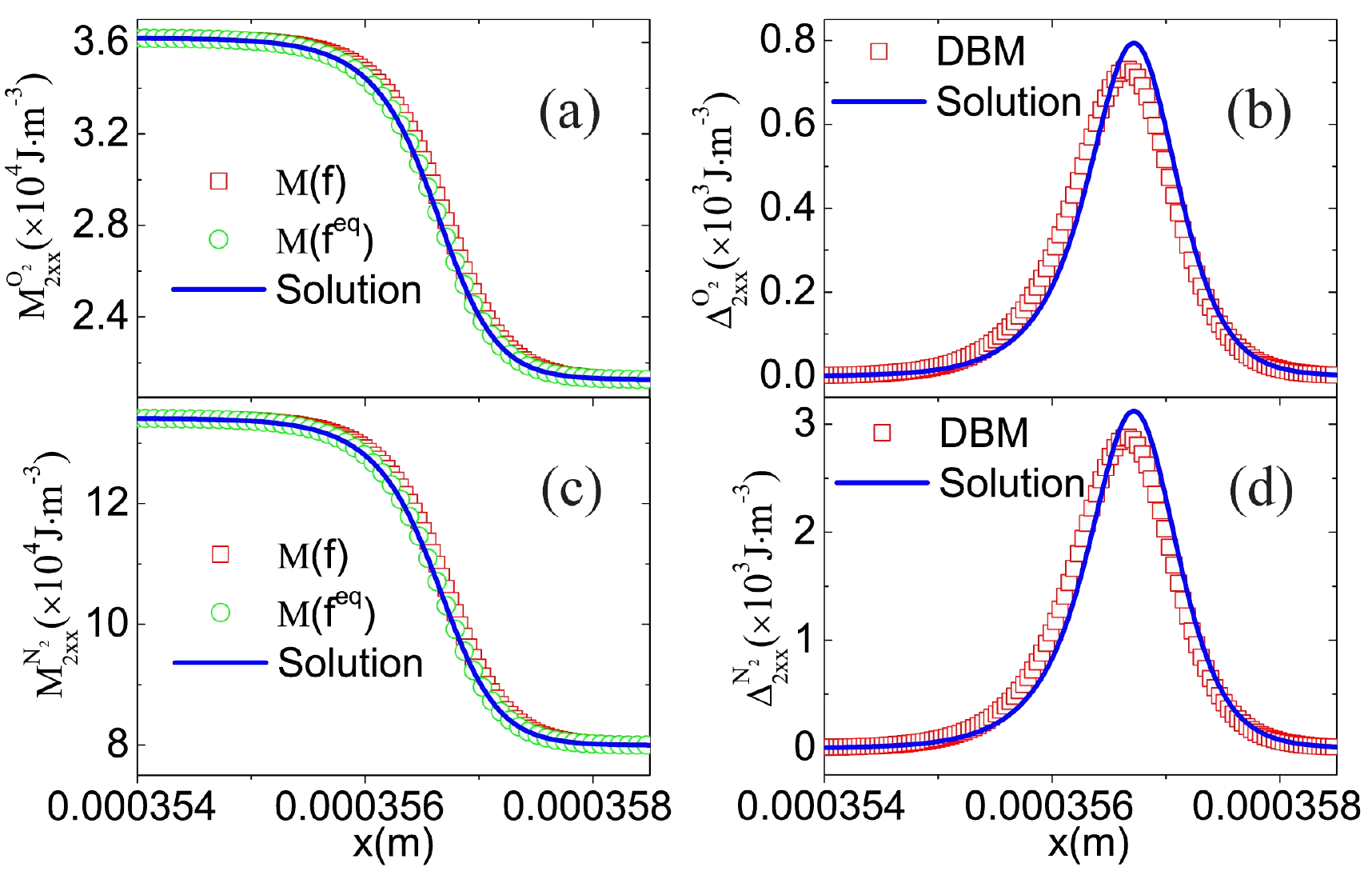}
	\end{center}
	\caption{Nonequilibrium manifestations around the shock wave: (a) translational energy of oxygen in $x$ direction, (b) departure of translational energy of oxygen in $x$ direction from equilibrium state, (c) translational energy of nitrogen in $x$ direction, (d) departure of translational energy of nitrogen in $x$ direction from equilibrium state.}
	\label{Fig06}
\end{figure}

A shock wave is a type of disturbance that propagates faster than the local speed of sound in a fluid with significant compressible effects. Its applications cover the fields of medicine, astrophysics, industrial engineering, etc. For example, it becomes effective medical treatment for kidney and ureteral stones. It can be used for cell transformation, preservative impregnation in bamboo, sandal oil extraction, and removal of micron size dust from silicon wafer surfaces \cite{Jagadeesh2008}. To validate the DBM for high-speed compressible systems, we conduct the simulation of a shock wave. The wave propagates in the air from left $L$ to right $R$. The initial field is, 
\begin{equation}
\left\{ \begin{array}{*{35}{l}}
{{\left( \rho ,{{u}_{x}},{{u}_{y}},T \right)}_{L}}=\left( 1.58407\text{kg}\cdot {{\text{m}}^{-3}}, 106.637\text{m}\cdot {{\text{s}}^{-1}}, 0\text{m}\cdot {{\text{s}}^{-1}}, 333.612\text{K} \right)  \\
{{\left( \rho ,{{u}_{x}},{{u}_{y}},T \right)}_{R}}=\left( 1.17092\text{kg}\cdot {{\text{m}}^{-3}}, 0\text{m}\cdot {{\text{s}}^{-1}}, 0\text{m}\cdot {{\text{s}}^{-1}}, 300\text{K} \right)  \\
\end{array} \right.
\tt{.}
\label{ConfigurationShock}
\end{equation}
The grid is $N_x \times N_y=40000 \times 1$, spatial step $\Delta x= \Delta y =1 \times 10^{-8}$ m, temporal step $\Delta t =1.25 \times 10^{-12}$ s. Figure \ref{Fig05} plots the profiles of the shock: (a) $\rho$, (b) $u_x$, (c) $T$, (d) $p$. The squares denote DBM results, the lines exact solutions. The DBM results behind the shock wave are $(\rho, u_x, u_y, T)=(1.58407\text{kg}\cdot {{\text{m}}^{-3}}, 106.637\text{m}\cdot {{\text{s}}^{-1}}, 0\text{m}\cdot {{\text{s}}^{-1}}, 333.612\text{K})$, which equal the exact values in Eq. (\ref{ConfigurationShock}) precisely. 

To exhibit the capability of the DBM to study nonequilibrium behaviours, Fig. \ref{Fig06} shows the nonequilibrium manifestations around the shock wave. Figure \ref{Fig06} (a) displays the translational energy of oxygen in $x$ direction $\frac{1}{2}\text{M}_{2xx}^{{{\text{O}}_{2}}}\left( f \right)$, its equilibrium counterpart $\frac{1}{2}\text{M}_{2xx}^{{{\text{O}}_{2}}}\left( f^{eq} \right)$, and the exact solution ${{n}^{{\text{O}}_{2}}}T+{{\rho }^{{\text{O}}_{2}}}u_{x}^{2}$. Figure \ref{Fig06} (b) illustrates the departure of translational energy of oxygen in $x$ degree of freedom from its equilibrium state $\frac{1}{2} \Delta_{2xx}^{\tt{O}_2 }$. Figures \ref{Fig06} (c) and (d) are for nitrogen. It is clear that $\text{M}_{2xx}^{\sigma}\left( f^{eq} \right)$ coincides with the solution ${{n}^{\sigma }}T+{{\rho }^{\sigma }}u_{x}^{2}$ in panels (a) and (c), respectively. Physically, the translational energy of oxygen (or nitrogen) in $x$ degree of freedom travels faster than its equilibrium counterpart. Consequently, its departure from the equilibrium state is greater than zero around the shock wave. Furthermore, there are few differences between the DBM and the approximate solution \cite{Lin2016CNF} in panels (b) and (d), respectively. Because the approximate solution is obtained by the first-order truncation of distribution function \cite{Lin2016CNF}. The simulation results are satisfactory. 

\subsection*{Supersonic reacting wave}

\begin{figure}[tbp]
	\begin{center}
		\includegraphics[width=0.8\textwidth]{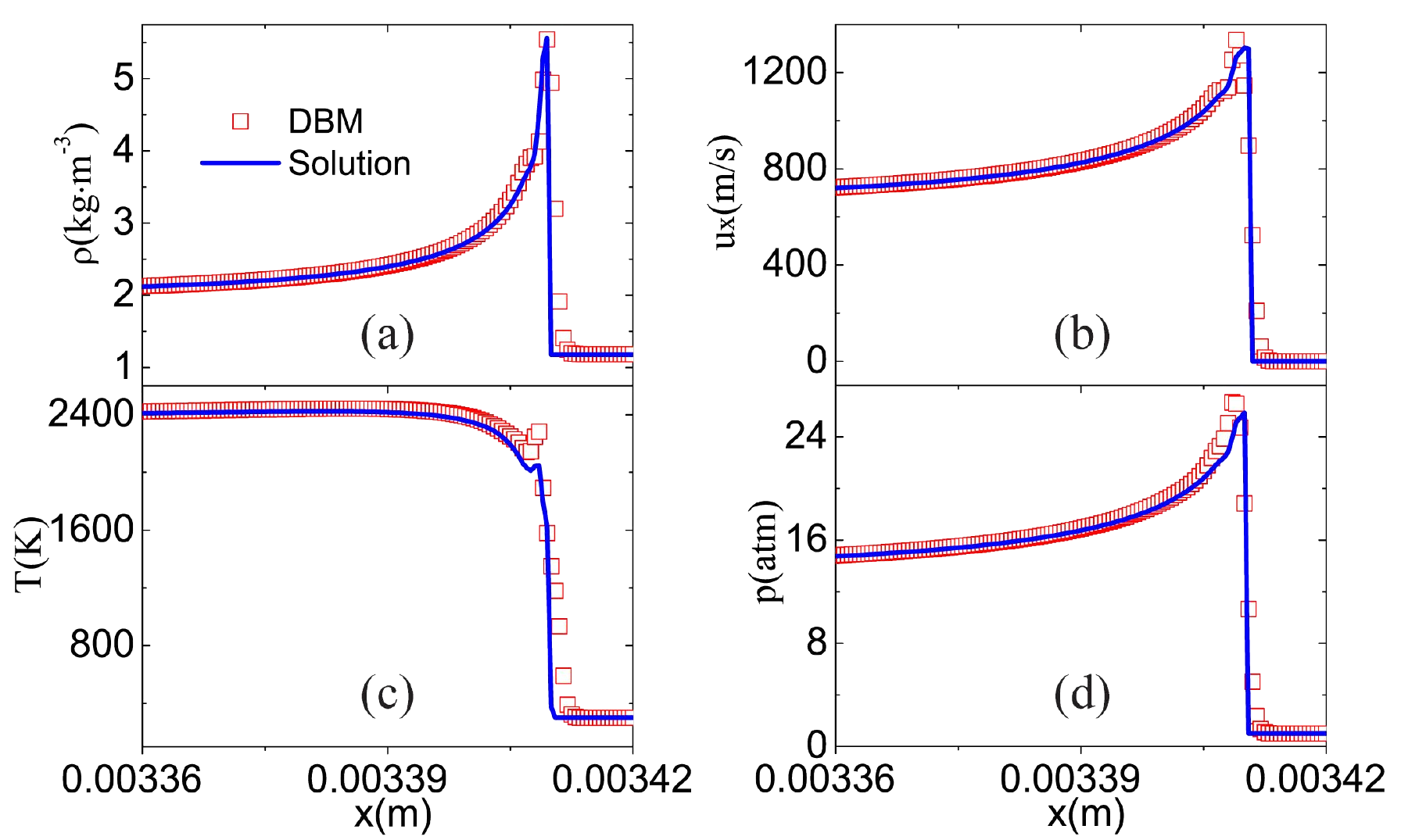}
	\end{center}
	\caption{Profiles of a supersonic reacting wave: (a) mass density $\rho$, (b) velocity $u_x$, (c) temperature $T$, (d) pressure $p$. The squares stand for DBM results, the lines exact solutions.}
	\label{Fig07}
\end{figure}

Supersonic reactive flows have been successfully used to deposite coating to a surface, clean equipment, mine for minerals, weld metals, etc. Numerical research of supersonic reacting wave has practical significance for the prevention of gas explosion in mining, flammable dust fires, and furnace burst in industry, etc. For the sake of verifying its suitability for supersonic reactive flows, the model is used to simulate a reacting wave. The initial field is divided into two parts. The right part is occupied by the premixed propane-air with equivalence ratio $0.524865$, the left part by the chemical products. The reacting wave travels from left to right. And physical quantities satisfy the Hugoniot relations, i.e.,
\begin{equation}
\left\{ \begin{array}{*{35}{l}}
{{\left( \rho ,{{u}_{x}},{{u}_{y}},T \right)}_{L}}=\left( 2.00166\text{kg}\cdot {{\text{m}}^{-3}}, 666.352\text{m}\cdot {{\text{s}}^{-1}}, 0\text{m}\cdot {{\text{s}}^{-1}}, 2363.81\text{K} \right)  \\
{{\left( \rho ,{{u}_{x}},{{u}_{y}},T \right)}_{R}}=\left( 1.18420\text{kg}\cdot {{\text{m}}^{-3}}, 0\text{m}\cdot {{\text{s}}^{-1}}, 0\text{m}\cdot {{\text{s}}^{-1}}, 300\text{K} \right)  \\
\end{array} \right.
\tt{.}
\label{ConfigurationDetonation}
\end{equation}
The grid is $N_x \times N_y=8000 \times 1$, spatial step $\Delta x= \Delta y = 5 \times 10^{-7}$ m, temporal step $\Delta t =6.25 \times 10^{-11}$ s. 

Figure \ref{Fig07} displays the wave profiles: (a) $\rho$, (b) $u_x$, (c) $T$, (d) $p$. The squares indicate DBM results, the lines analytic solutions of Zeldovich-Neumann-Doering (ZND) theory \cite{Law2006}. The DBM results behind the wave are $(\rho, u_x, u_y, T)=(2.00166\text{kg}\cdot {{\text{m}}^{-3}}, 666.356\text{m}\cdot {{\text{s}}^{-1}}, 0\text{m}\cdot {{\text{s}}^{-1}}, 2383.86\text{K})$. Compared with the first row in Eq. (\ref{ConfigurationDetonation}), the relative differences are $0\%$, $0\%$, $0\%$, and $0.8\%$, respectively. Moreover, DBM gives the wave speed $1632 \text{m}/\text{s}$, and the analytic solution is $1631.6 \text{m}/\text{s}$. The relative difference between them is $0.02\%$. Additionally, there are slight differences between the theoretical and numerical results around the wave peak. Physically, the ZND theory assumes a sharp discontinuity at the wave peak and ignores the viscosity, heat conduction and other nonequilibrium effects \cite{Law2006}. On the other hand, the DBM takes into account the viscosity, heat conduction and other transport processes. Thus, the DBM is more reliable than the simple ZND theory. 

\subsection*{Kelvin-Helmholtz instability}

\begin{figure}[tbp]
	\begin{center}
		\includegraphics[width=0.8\textwidth]{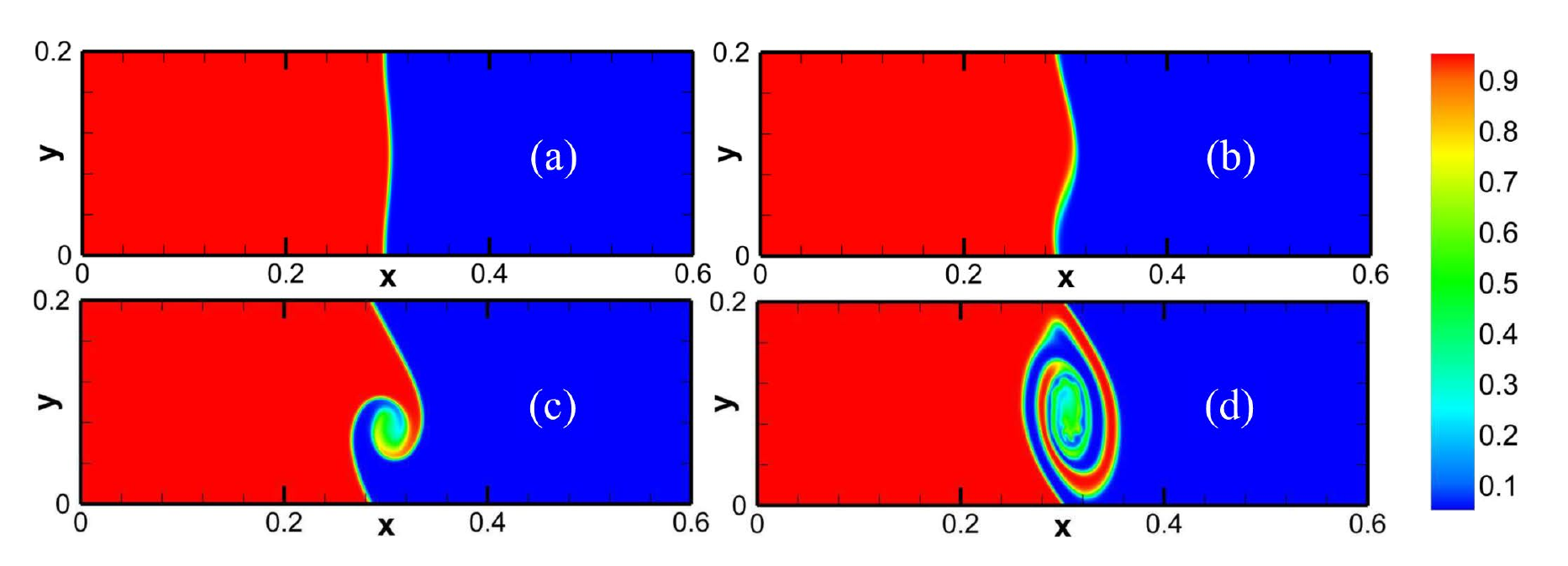}
	\end{center}
	\caption{The molar fraction of propane in the evolution of KH instability at various times: (a) $0$ s, (b) $5\times 10^{-4}$ s, (c) $10^{-3}$ s, (d) $2\times 10^{-3}$ s.}
	\label{Fig08}
\end{figure}
\begin{figure}[tbp]
	\begin{center}
		\includegraphics[width=0.5\textwidth]{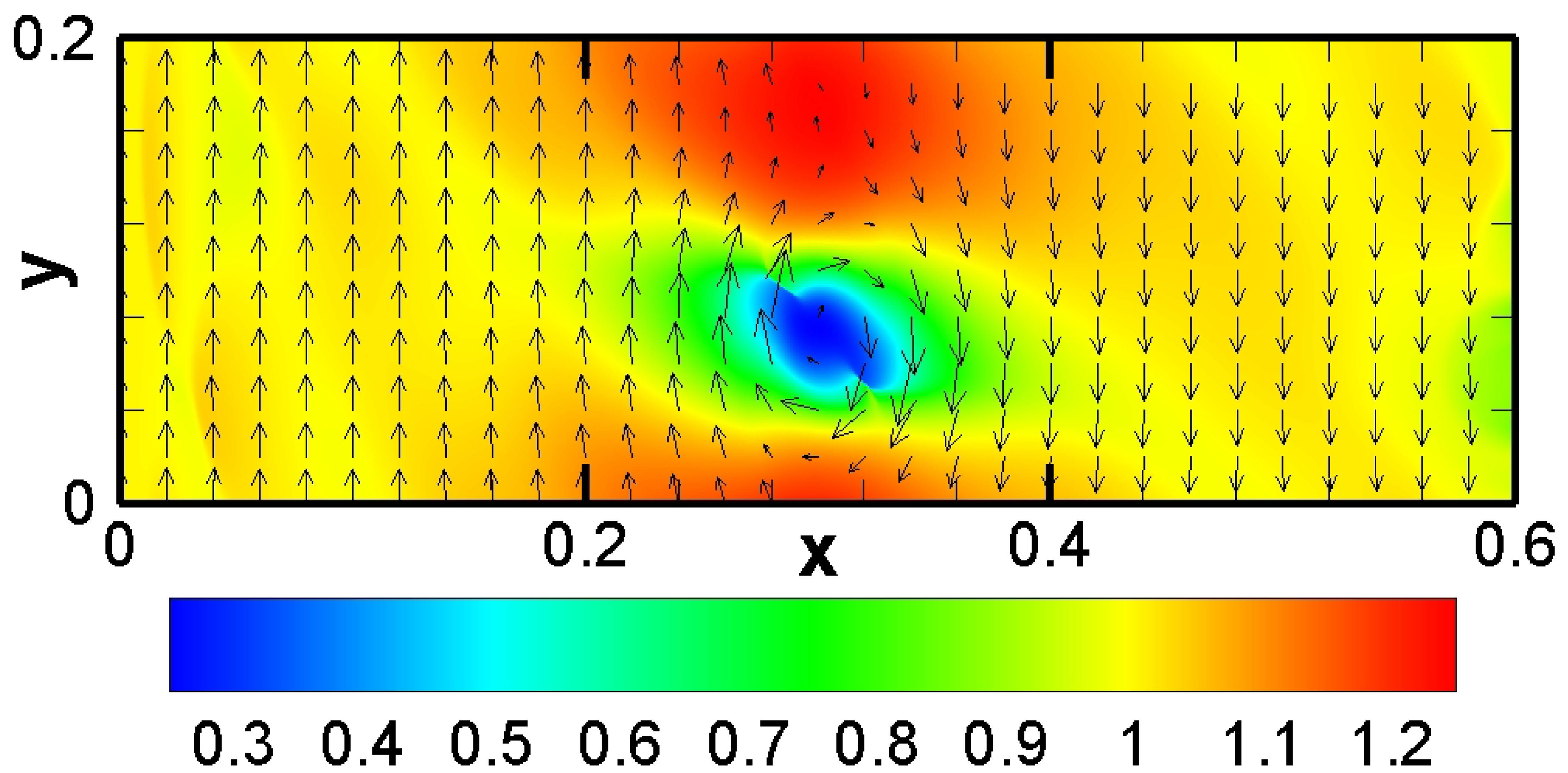}
	\end{center}
	\caption{Pressure and velocity fields in the evolution of KH instability at time $10^{-3}$ s.}
	\label{Fig09}
\end{figure}
\begin{figure}[tbp]
	\begin{center}
		\includegraphics[width=0.45\textwidth]{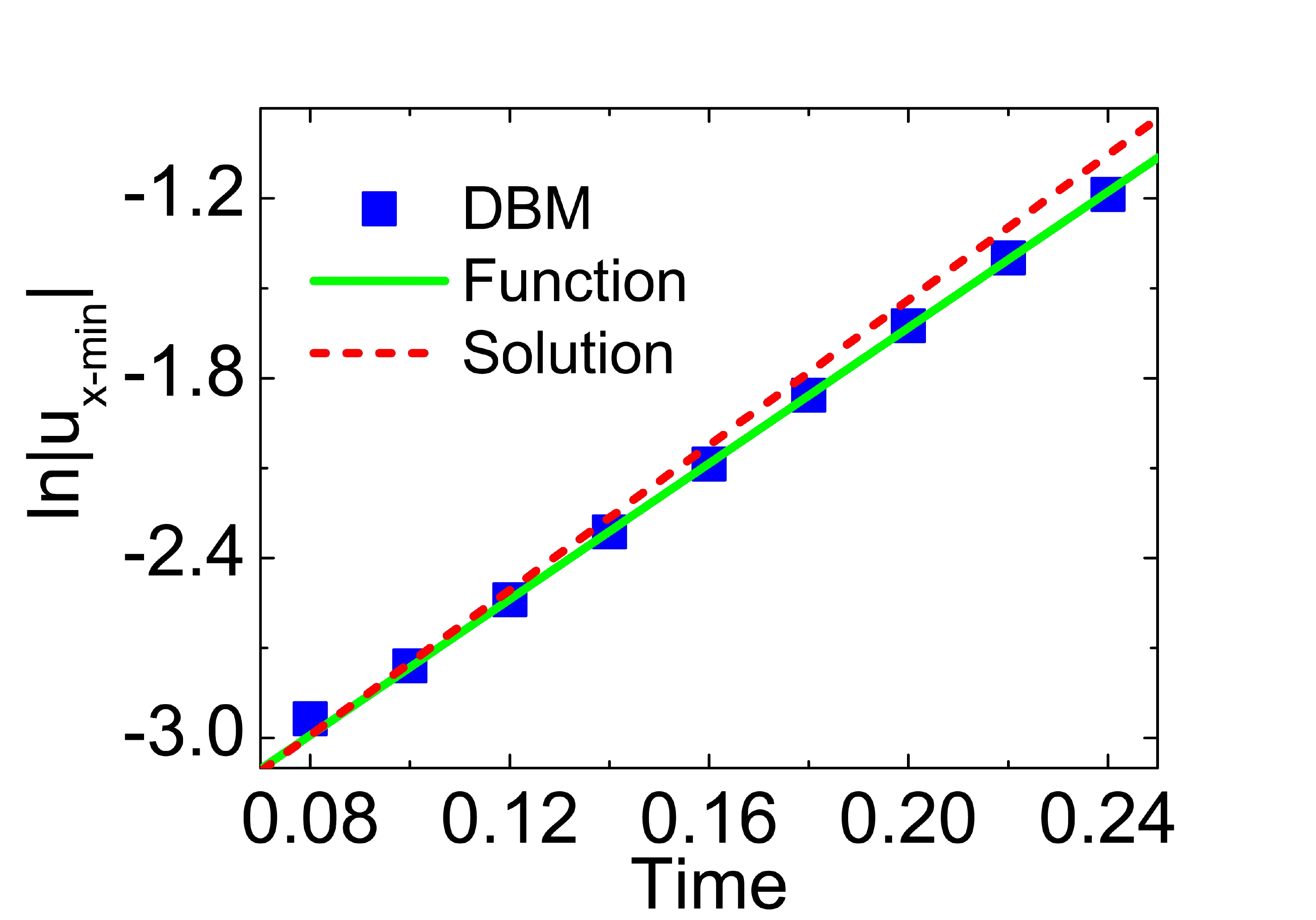}
	\end{center}
	\caption{Evolution of the velocity perturbation, $\tt{ln}|u_{x-min}|$. The squares stand for DBM results, the solid (dashed) line is for the fitting function (analytic solution).}
	\label{Fig10}
\end{figure}

To demonstrate that the DBM has a good ability of capturing interface deformation, we simulate a typical 2-D phenomenon, KH instability. The initial field, with area $0.6 \tt{m} \times 0.2\tt{m}$, consists of two parts. The left part is full of propane with vertical speed $200 \tt{m} \cdot \tt{s}^{-1}$, while the right part is filled with air with $-200 \tt{m} \cdot \tt{s}^{-1}$. Considering the transition layer between the two parts, the field jump at the interface is smoothed by a tanh profile with width $0.002 \tt{m}$. The temperature is $300$ K, and pressure $1$ atm. Between the propane and air is a sinusoidal interface with amplitude $0.003 \tt{m}$, which is used to promote the KH instability.  Moreover, the outflow and periodic boundary conditions are adopted in the $x$ and $y$ directions, respectively. The grid is $N_x \times N_y=3000 \times 1000$, spatial step $\Delta x= \Delta y =2 \times 10^{-4}$ m, temporal step $\Delta t =2.5 \times 10^{-8}$ s. 

Figure \ref{Fig08} shows the molar fraction of propane at four different times. Initially, the interface starts to wrinkle due to the initial perturbation and velocity shear. A rolled-up vortex emerges after the initial linear growth stage. Then there is a large vortex around the interface. The evolution of the field is qualitatively similar to previous studies \cite{Lin2016CNF,XuGan2011}. Moreover, Fig. \ref{Fig09} delineates the contour of pressure with velocity field, corresponding to Fig. \ref{Fig08} (c). Compared Fig. \ref{Fig09} with Fig. \ref{Fig08} (c), we can find that the minimum pressure, $p=0.257\tt{atm}$, is located at, ($0.3004 \tt{m}$, $0.0750 \tt{m}$), the center of the vortex. While the maximum, $p=1.24\tt{atm}$, takes place at, ($0.3008 \tt{m}$, $0.1708 \tt{m}$), where counterflows above the vortex encounter each other and the horizontal velocity is close to zero. Physically, the pressure gradient around the vortex provides the centripetal force for the rotating flow. 

To quantitatively validate the results, we plot the logarithm of absolute value of the minimum perturbed horizontal velocity, $\tt{ln}|u_{x-min}|$, versus time, see Fig. \ref{Fig10}. The squares are for DBM results, the solid line represent the fitting function $F(t)=-3.89713 + 11.3302 t$, and the dashed line stands for the analytic solution $F(t)=-3.95868 + \dot{A} t$ with the growth rate $\dot{A}=12.0995$ \cite{Wang2010}. Here the nondimensionalization is used the same as Ref. \cite{Wang2010}. The relative difference of the growth rate between the fitting function and the analytic solution is $6.4\%$. Furthermore, we compare the simulation frequency $1256$ Hz with the analytic solution $1248$ Hz \cite{Lin2016CNF,Wang2010}. The relative difference is $0.6\%$. The difference mainly comes from the fact that the effects of compressibility, viscosity, and heat conduction are considered by the DBM, but ignored by the analytic theory \cite{Wang2010}. 

\section*{Conclusions}
We present a reactive multi-component DBM in combination with a one-step overall chemical reaction. The effects of chemical reaction and external force are considered. A two-dimensional sixteen-velocity model D2V16 is proposed with adjustable parameters ($v_{a}^{\sigma }$, $v_{b}^{\sigma }$, $v_{c}^{\sigma }$, $v_{d}^{\sigma }$) controlling discrete velocities and (${\eta }_{a}^{\sigma }$, ${\eta }_{b}^{\sigma }$, ${\eta }_{c}^{\sigma }$, ${\eta }_{d}^{\sigma }$) for internal energies in extra degrees of freedom. The specific heat ratio of each species $\sigma$ is flexible since extra degrees of freedom are taken into account. This model is suitable for premixed, nonpremixed or partially premixed combustion, from subsonic to supersonic fluid flows, in or out of equilibrium. Through the Chapman-Enskog multiscale analysis, the DBM recovers the modified NS equations for reactive species with external force effects in the hydrodynamic limit. In addition to the usual nonequilibrium terms (viscous stress and heat flux) in NS models, more detailed hydrodynamic and thermodynamic nonequilibrium quantities (high order kinetic moments and their departure from equilibrium) can be calculated in the DBM dynamically and conveniently. 
Since the DBM can provide detailed distributions of nonequilibrium quantities, it permits to assess the corresponding numerical predictions of NS models without considering the nonequilibrium effects.
Hence, the DBM has the potential to offer more accurate information to help design devices operating in transient and/or extreme conditions away from equilibrium. Furthermore, thanks to its mesoscopic nature, the DBM could provide deeper insight into ubiquitous reactive or nonreactive fluid flows with a large span of spatial-temporal scales. Finally, due to its generality, the developed methodology is applicable to a wide range of phenomena across many energy technologies, emissions reduction, environmental protection, mining accident prevention, chemical and process industry. 


\section*{Acknowledgements}

The authors thank Profs. Aiguo Xu and Guangcai Zhang for their helpful suggestions. This work is supported by the Natural Science Foundation of China (NSFC) under Grant No. 91441120 and the Center for Combustion Energy at Tsinghua University. Support from the UK Engineering and Physical Sciences Research Council under the project ``UK Consortium on Mesoscale Engineering Sciences (UKCOMES)'' (Grant No. EP/L00030X/1) is gratefully acknowledged.

\section*{Author contributions}

S.S. contributed to the analysis of the theory and results of the DBM; L.F. provided the code of standand LBMs and assisted in performing the simulations; K.L. initiated and supervised the study; C.L. developed and validated the DBM as well as drafted the manuscript. All authors modified and approved the manuscript. 

\section*{Additional information}

\textbf{Competing financial interests}: The authors declare no competing financial interests.

\end{document}